\documentclass[referee]{aa}
\usepackage{graphicx}
\usepackage{txfonts}
\usepackage{epstopdf}

\usepackage{booktabs}

\def\aap{A\&A }                % Astronomy and Astrophysics
\def\Lunar and Planetary Science Conference 
Proceedings{Proc.~Lunar~Sci.~Conf. 
} 
\def\Lunar and Planetary Institute Science Conference 
Abstracts{Lunar~Planet.~Sci. }    % Journal of Geophysics Research
\def\apjl{Astronomical Physical Journal letters }

\begin{document}

   \title{Composition of Jupiter irregular satellites sheds light on their 
origin}

   %\subtitle{I. Overviewing the $\kappa$-mechanism}

   \author{M. Bhatt
          \inst{1}
          \and
          V. Reddy\inst{2}
          \and
          K. Schindler\inst{3,4}
          \and
          E. Cloutis\inst{5}
          \and
          A. Bhardwaj\inst{6}
          \and
          L. Le Corre\inst{7}
          \and
          P. Mann \inst{5}
          }

   \institute{Space Physics laboratory, Vikram Sarabhai Space Centre, 
              Thiruvananthapuram, 695022, Kerala, India\\
              \email{mu$\_$bhatt@vssc.gov.in}
         \and
             Lunar and Planetary Laboratory, University of Arizona, 1629 E,
             University Blvd, Tucson, AZ 85721-0092\\
             \email{reddy@lpl.arizona.edu}
         \and
        Deutsches SOFIA Institut, Universität Stuttgart, Pfaffenwaldring 29, 70569 Stuttgart, Germany
\and SOFIA Science Center, NASA Ames Research Center, Mail Stop N211-1, Mo ett Field, CA 94035, USA\\
         \email{schindler@dsi.uni-stuttgart.de}
         \and
         Department of Geography, University of Winnipeg, 515 Portage 
         Avenue, Winnipeg, Manitoba, Canada R3B 2E9\\ 
         \email{e.cloutis@uwinnipeg.ca, paul.mann347@gmail.com}
         \and
         Physical Research laboratory, Ahmedabad, 380009, Gujarat, India\\
         \email{abhardwaj@prl.res.in}
         \and
         Planetary Science Institute, 1700 East Fort Lowell Road, Tucson, 
         AZ, 85719, USA\\
         \email{lecorre@psi.edu}
             }
   \date{\today}

% \abstract{}{}{}{}{} 
% 5 {} token are mandatory

  \abstract
   {Irregular satellites of Jupiter  with  their  highly eccentric, inclined and 
distant orbits suggest that their capture took place just before the giant 
planet migration.}
   {We aim to improve our understanding of the surface composition of irregular satellites of Jupiter to gain 
insight into a narrow time window when our Solar System was forming.}
   {We observed three Jovian irregular satellites, Himalia (JVI), Elara (JVII), and 
Carme (JXI), using a medium-resolution 0.8-5.5~$\mu$m spectrograph, 
SpeX  on the National Aeronautics and Space Administration (NASA) Infrared 
Teselscope Facility (IRTF). Using a linear spectral unmixing 
model we have constrained the major mineral phases on the surface of these 
three bodies.}
   {Our results confirm that the surface of Himalia (JVI), Elara (JVII), and 
Carme (JXI) are dominated by opaque materials such as those seen in 
carbonaceous chondrite meteorites. Our spectral modeling of NIR spectra 
of Himalia and Elara confirm that their surface composition is the same 
and magnetite is the dominant mineral. A comparison of the spectral shape of 
Himalia with the two large main C-type 
asteroids, Themis (D$\sim$176~km) and Europa (D$\sim$352~km), suggests  
surface composition similar to Europa. The NIR spectrum of Carme exhibits blue 
slope up to 
1.5~$\mu$m and is spectrally distinct from those of Himalia and Elara. Our 
model suggests that it is compositionally similar to amorphous carbon.}
{Himalia and Elara are compositionally similar but differ significantly from 
Carme. These results support the hypotheses that the Jupiter's irregular 
satellites are captured bodies that were subject to further breakup events 
and clustered as families based on their similar 
physical and surface compositions.}
   \keywords{Planets and satellites: individual: Himalia, Elara, Carme --
surfaces --
Techniques: spectroscopic}
%\authorrunning{M. Bhatt et al.}
   \maketitle
%
%-------------------------------------------------------------------

\section{Introduction} \label{sec:intro}

The Jovian planets of our solar system have both regular and irregular moons, 
which are grouped based on their orbital characteristics. 
Regular satellites are on prograde orbits (inclinations $<90^{\circ}$) with low 
eccentricities and low orbital inclinations suggesting that they formed  in a 
planetocentric disk depending on thermal and chemical conditions in the 
protoplanetary nebulae \citep{1979Pollack}. The irregular satellites, 
however, are smaller in size and their orbits are highly inclined and 
eccentric suggesting a distinct origin mechanism.  Two competing hypotheses have 
been proposed  to explain the origin of these irregular satellites. The first 
is aerodynamic capture by Jupiter of irregular satellites from heliocentric 
orbits \citep{1977Heppen,1979Pollack} or by disruptive collisions 
\citep{1971Colombo}. Recent works invoke the capture 
of these irregular satellites from the outer solar system after the giant 
planets' migration \citep{2007Nesvorny,2014Nersvorny}. These captured objects 
are important to study as they may represent the planetesimals that coalesced 
and formed the core of the giant planets \citep{2008Nicholson}. Furthermore, 
the surface composition of the irregular satellites can be used to understand 
their origin and the conditions under which they formed in the solar 
nebula. 

The irregular satellites of giant planets have been further divided into 
prograde (inclinations $<90^{\circ}$) and retrograde groups (inclinations 
$>90^{\circ}$) based on their orbital properties. These groups  are further 
clustered  into families according to their orbital period. \cite{1987Hartmann} 
and \cite{2003Grav} used optical measurements for a larger set of irregular 
satellites, and found that the colour variations are homogeneous  within a 
family, indicating that each retrograde and prograde family is the result of 
the capture of a single parent body followed by breakup events 
\cite[e.g.][]{1971Bailey,1971BaileyJGR,1979Pollack,2001Rettig, 
2003Grav,2003sheppard,2004Grav,2007Nesvorny,2007Jewitt}. Jupiter is known to 
have at least five such families, named after the largest member of each 
family, two of which (Himalia and Carpo) are prograde  and the  other three 
(Ananke, Carme, and Pasipha\"e) are retrograde \citep{2003Nesvorn,2003sheppard}. 
Here we present low-resolution telescopic spectra of three of Jupiter's 
irregular satellites, Himalia, Elara, and Carme, in an effort to identify 
spectrally dominant minerals on their surfaces. These three objects have been 
selected based on the favourable observing conditions around two Jupiter 
oppositions in 2012 and 2013, as described in Table~\ref{tab_obs}.

Himalia and Elara both belong to the Himalia dynamical family of the prograde 
group that orbits Jupiter at distances of about 11.5 million~km ($\sim$165 
Jovian radii), with orbital periods of 250.6 days and 259.7 days, respectively 
\citep{2000Jacobson}. Carme belongs to the retrograde group and orbits Jupiter 
at a distance of about 23.4 million~km ($\sim$334 Jovian radii) with an orbital 
period of 702.2 days \citep{2000Jacobson}. Himalia is known to be an 
elongated object with axes of 150x120~km based on Cassini spacecraft 
measurements  \citep{2003Porco}.  Recent studies by \cite{2015Grav} 
suggest that Himalia has low thermal inertia, with significant surface 
roughness, and a visible albedo of 5.7$\pm$0.8\%. Elara is comparatively small 
with a diameter of $\sim$80~km.   The radiometric observations by 
\cite{1977Cruikshank} derived an albedo of 3$\pm$1\% for Elara. This value was 
refined to 4.6$\pm$0.7\%  by \cite{2015Grav} who reported a low beaming value of 
0.79$\pm$0.03 indicating significant surface roughness. Carme, with a diameter 
of $\sim$50~km, is the biggest member of the Carme family, a dynamical family 
of 19 known irregular satellites. \cite{2015Grav} estimated its albedo to be 
3.5$\pm$0.6\%. 

The large Jovian irregular satellites have been observed and analyzed 
using their light curves, colours, and reflectance spectra but the reported 
measurements are sometimes contradictory. The multicolour observation of some 
retrograde and prograde Jovian irregular satellites by \cite{1984Tholen} 
suggested C-class surface composition for  prograde and more diverse colours for 
the retrograde families with a mixture of C- and D-type spectra. They noted 
that Carme had a  flat visible wavelength reflectance spectrum, but with a 
strong upturn in the ultraviolet. \cite{1984Tholen} suggested that  Carme might 
be showing low-level cometary activity with CN emission at 0.388~$\mu$m.
\cite{1991Luu} identified C-and D-type asteroid spectral features for  both 
prograde and retrograde families based on spectroscopic observations of JV-JXIII 
and suggested them to be similar to Jupiter's Trojan asteroids. Based on 
1.3-2.4~$\mu$m near-IR observations, \cite{2000Brown} reported their 
compositions as being similar to P- and D-class asteroids from the outer asteroid 
belt, while their visible spectra resemble C-class asteroids. 
\cite{1977Cruikshank} and \cite{1980Degewij} observed Himalia 
in the near-IR and confirmed that its surface composition is similar to that of 
C-type asteroids. \cite{2000Brown} concluded that NIR spectra of Himalia and Elara are featureless 
between 1.4 and 2.5~$\mu$m and do not contain any water-ice absorption features. Subsequently, \cite{2014Brown}  supported these findings and suggested that these objects lacked aqueously altered phyllosilicates based on the absence of a 3~$\mu$m absorption band. \cite{2003Brown} and \cite{2004Chamberlain} studied Himalia using 
data acquired by the Visual and Infrared Mapping Spectrometer (VIMS) on-board 
Cassini spacecraft during Jupiter's fly-by and found that its spectrum 
(0.3-5.1~$\mu$m) has low reflectance, a slight red slope, and an apparent 
absorption near 3-$\mu$m suggesting the presence of water in some form. 
In addition,  \cite{2000Jarvis} reported a weak absorption at 
0.7~$\mu$m in Himalia's spectrum and attributed it to oxidized iron. 
Contrary to this result, \cite{2014Brown} found no evidence for aqueously 
altered phyllosilicates  in the 2.2-3.8~$\mu$m region.

\section{Observation and data reduction} \label{sec:obs}

Spectral observations of Himalia (JVI), Elara (JVII) and Carme (JXI) were 
obtained remotely using a  medium-resolution 0.8-5.5~$\mu$m 
spectrograph, SpeX  on the National Aeronautics and Space Administration (NASA) 
Infrared Teselscope Facility (IRTF). The instrument was operated in 
low-resolution (R 150) prism 
mode with a 0.8" slit. Observational circumstances for the three objects are 
listed 
in Table~\ref{tab_obs}. Local G-type stars were observed before and after the 
satellites in order to correct 
for telluric features. Observing conditions were less than ideal (high cirrus) 
when we observed Elara on Dec. 12 2012, 
and hence its spectrum has lower SNR than Carme despite Elara being 1.3 
magnitudes brighter. The Solar analogue star SAO93936 was also observed and used to 
correct for spectral slope variations that could arise due to the use of a 
non-solar extinction star. The slit was oriented along the parallactic angle in 
order to reduce the effects of differential atmospheric 
refraction during the observational period. 

\begin{figure}
\centering
\includegraphics[height=0.6\textheight]{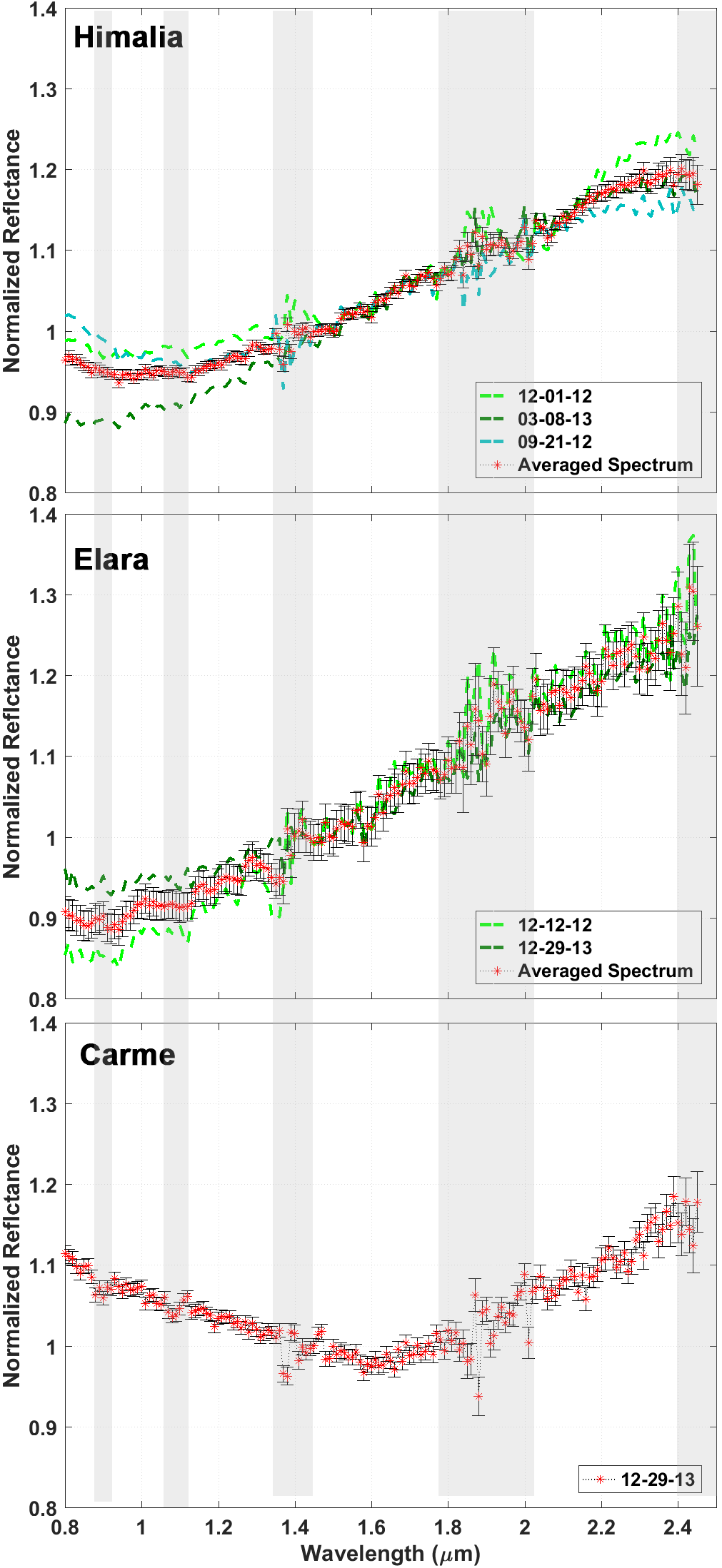}
\caption{ Normalised  near-IR reflectance spectra of Jupiter irregular 
satellites, Himalia, Elara and Carme, obtained using the SpeX instrument on the 
NASA IRTF. The spectra are normalised to unity at 1.5~$\mu$m. The 
averaged spectrum used for linear spectral unmixing is also shown. 
Observational 
circumstances are listed in Table~\ref{tab_obs}. 
Higher scatter in some wavelength ranges is due to incomplete telluric 
correction. The grey bands indicate the location of the telluric bands.}
\label{fig_HEC_R}
\end{figure}

The prism data were processed using 
the IDL-based Spextool provided by the NASA IRTF \citep{2004Cushing}.  A 
detailed description of the steps 
followed in the data reduction process can be found in 
\cite{2013Sanchez,2015Sanchez}. Figure~\ref{fig_HEC_R}  shows the relative 
flux spectra of Himalia, Elara, and Carme normalised at 
1.5~$\mu$m. Each spectrum plotted in Fig.~\ref{fig_HEC_R} is an 
average of all NIR spectra observed on a single night. The 
average spectrum have been obtained using Spextool.  Higher data scatter for 
the wavelength region around 1.9~$\mu$m and beyond 2.4~$\mu$m is due to the 
incomplete telluric correction (grey bars on Fig.~\ref{fig_HEC_R}). The greater 
scattering for a specific wavelength region can 
introduce errors in spectral unmixing analysis. Therefore, we restricted the 
spectral unmixing analysis to wavelengths between 0.8 and 2.40~$\mu$m  with a wavelength interval of 
0.01~$\mu$m. We used the average spectrum for modeling the surface composition. 
Nightly average spectra for Himalia on three different epochs show a broad 
absorption feature at 1~$\mu$m and a moderately red slope between 1.4 and 
2.5~$\mu$m. Similarly, Elara was observed on two nights and its spectra 
show a moderately 
red slope between 1.4 and 2.5~$\mu$m but with a shallower absorption band 
at 1~$\mu$m compared to Himalia. Carme was only observed for one night and its 
spectrum shows an absorption band at 1.6~$\mu$m that is broader than the other 
two objects.  

%%%%%%%%%%%%%%%%Table%%%%%%%%%%%%%%%%%%%%%%%%
\begin{table*}
%\begin{table}
\caption{The observational circumstances.}
\label{tab_obs}
\begin{center}
\begin{tiny}
\begin{tabular}{ c c c c c c c c c}

\hline\hline
Object& Date of observation& Start UTC  & End UTC&V.Mag &Phase Angle &Airmass 
&Local standard star&Solar analogue\\\midrule
Himalia (JVI)   & 09/21/2012    & 14:04 & 14:44 & 14.96 & 11.40 &1.00-1.04 & 
GSC 01293-00022 & SAO93936\\
                & 12/01/2012    & 11.10 & 11.45 & 14.76 & 0.35 & 1.02-1.07 & HD 
29714 & SAO93936\\
                & 03/08/2013    & 5:10  & 5:50  &15.9   &11.20  &1.06-1.10 & HD 
284507 &SAO93936\\
Elara (JVII)    & 12/12/2012    & 10:50 & 11:36 & 16.62 & 2.26 & 1.06-1.10 & 
SAO76633 & SAO93936\\
                & 12/29/2013    & 08:44 & 09:20 & 14.78 & 2.05 & 1.08-1.17 & 
GSC 01896-01098 & SAO93936\\
Carme (JXI)     & 12/29/2013    & 12:27 & 12:57 & 17.89 & 2.03 & 
1.05-1.10 & GSC 01358-02213 & SAO93936\\\hline

\end{tabular}
\end{tiny}
\end{center}
\end{table*}

%%%%%%%%%%%%%%%%%Figure%%%%%%%%%%%%%%%%%%%%%%%%
%%%%%%%%%%%%%%%%%Figure%%%%%%%%%%%%%%%%%%%%%%%%
\section{Modeling strategy: Linear spectral unmixing} \label{sec:model}

Several spectral mixing algorithms have been proposed for studying the surface 
of 
various planetary bodies  
\cite[e.g.,][]{1986Adams,1990Sunshine,2000van,2002Hapke,Shkuratov1999a,
2009Poulet,2014Horgan,2016Marsset,2016Zambon}. 
 Mixing models are specifically useful for determining mineral abundances when 
laboratory spectral calibrations are not available.  Mixing models are 
complimentary to spectral parameter studies using absorption band centre 
because it allows 
for the mathematical description of the band shapes and the qualitative and 
quantitative assessment of the absorption features. However,  in general, models 
do not provide a unique solution but a set of possible 
solutions. The observed reflectance spectra from any planetary body are mostly 
a non-linear mixture of spectra from different minerals due to multiple 
interactions of  the incidence light before being reflected back in the 
direction of 
the detector. In many cases, modeling of these reflectance spectra has been 
successfully accomplished using linear and non-linear unmixing models with 
certain uncertainties.  The non-linear models 
are more accurate than the linear models, however, they require additional 
information on scattering parameters, optical constants, mineral assemblages, 
and grain sizes \cite[e.g.,][]{1981Hapke,Shkuratov1999a}.  Linear unmixing, on 
the other hand, can be used to describe areal mixtures of surface materials by 
linear combinations of spectral components  
\cite[e.g.,][]{1986Adams,2008Combe,2016Zambon}. A limitation of this approach 
is that it is strictly valid only when the minerals are arranged
in discrete patches at the surface, which is certainly not
the case for all planetary surfaces. 

In our study, we used a linear unmixing algorithm that has been successfully 
applied to several planetary bodies to constrain their surface 
composition 
\cite[e.g.][]{2008Combe,2009Poulet,2014Horgan,2016Marsset,
2016Zambon}. Our goal is not to estimate the mineral distribution, but  to 
constrain the contribution of major minerals to the spectrum. We selected a 
classical Spectral Mixture Analysis (SMA) approach.  Our assumption 
is that the reflectance spectra of these irregular bodies is a mixture of a 
linear combination of a maximum of four components, selected from a large pool 
of laboratory data sets.  The SMA unmixing model is based on a least square 
inversion and can be represented in a standard way of  a linear combination 
of end members: 

 %%%%%%%%%%%%%%%%Equation%%%%%%%%%%%%%%%%%%%%%%%%
\begin{equation}
%\begin{align}
\label{eq_model}
Y=a1x1+a2x2+a3x3+...a_nx_n
%\end{align}
,\end{equation}
%%%%%%%%%%%%%%%%Equation%%%%%%%%%%%%%%%%%%%%%%%%
where $Y$ is the observed spectrum, x1 to x$_{n}$ are the normalised
laboratory spectra, and a1 to a$_{n}$ are their respective mixing 
coefficients. We constrain the algorithm output by fixing the additional 
condition that all the mixing coefficients should be positive and  their 
total contribution should be equal to unity \citep{1998Ramsey}.

The solution of Eq.~\ref{eq_model} is as follows:
 %%%%%%%%%%%%%%%%Equation%%%%%%%%%%%%%%%%%%%%%%%%
\begin{equation}
%\begin{align}
\label{eq_model_sl}
\mathrm{A} = (X^tX)^{-1}X^tY
%\end{align}
,\end{equation}
%%%%%%%%%%%%%%%%Equation%%%%%%%%%%%%%%%%%%%%%%%%

where A is a vector containing the mixing coefficients of the reference 
spectra (a1 to a$_{n}$) and X is a $NXM$ matrix of the end members (with 
N being the number of end members and M the number of spectral channels).

We included the option of automatic selection of multiple end members and 
evaluated the model fit by computing the coefficient of determination $R^2$. We 
selected a combination of three end members as an initial case and 
allowed a fourth reference spectrum in case the fit was not optimal. We did not 
observe any significant improvement in $R^2$ value when allowing a fifth or 
sixth reference spectrum in the model. We also observed that setting the number 
of end members to two decreases the quality of the model  and the model 
becomes unstable.

The coefficient vector A derived from the linear spectral unmixing model does 
not represent true mineral proportions in a rock, but rather indicates the 
major minerals with a qualitative estimate of their presence on the surface. We 
used  the simplex-projection unmixing algorithm from \cite{2011Heylen} for 
computing the coefficient vector A.

The model gives the best match based on the best $R^2$ value obtained. We 
considered the results of the ten best model runs out of all possible 
combinations of three or four end members from Table~\ref{tab_lab}. The model 
is 
considered stable  if minor changes in $R^2$ do not change the used
end members significantly. 

\subsection{End members selection} 
\label{sec:model_em}
%%%%%%%%%%%%%%%%Table%%%%%%%%%%%%%%%%%%%%%%%%
\begin{table}
\caption{End members of the selected sample library.}
\label{tab_lab}
\begin{center}
%\begin{tiny}
%\vspace{5pt}
\begin{tabular}{ c l l}
\hline\hline

S.N.    & Sample id     & Sample Type\\\midrule
1.      & LCA101        & Amorphous Carbon      \\
2.      & GRP101        & Graphite      \\
3.      & GRP102        & Amorphous Graphite    \\
4.      & MAG101        & Magnetite     \\
5.      & MAG102        & Magnetite     \\
6.      & ILM101        & Ilmenite\\
7.      & BER101        & Berthierine\\
8.      & CHM101        & Chamosite\\
9.      & CRO102        & Cronstedtite\\
10.     & GRE001        & Greenalite\\
11.     & ILL101        & Illite\\
12.     & MIN003        & Minnesotaite\\
13.     & MON102        & Montmorillonite\\
14.     & NON101        & Nontronite\\
15.     & SAP102        & Saponite\\
16.     & SRP104        & Serpentinite\\
17.     & SRP105        & Serpentinite\\
18.     & TOC101        & Tochilinite\\
19.     & TRO202        & Troilite\\
20.     & OOHOO2        & Bauxite\\
21.     & OOHOO4        & Gibbsite\\
22.     & OLV002        & Chrysolite\\
23.     & OLV106        & Forsterite\\
24.     & PYX003        & Bronzite\\
25.     & PYX036        & Augite\\
26.     & PLG103        & Albite\\
27.     & PLG125        & Plagioclase\\
28.     & SPI101        & Spinel\\
29.     & SPI127        & Spinel\\
30.     & CHR103        & Chromite\\
\hline
\end{tabular}
%\end{tiny}
\end{center}
\end{table}

%%%%%%%%%%%%%%%%Table%%%%%%%%%%%%%%%%%%%%%%%%

The selection of suitable end members is the most challenging and 
important step in the linear unmixing modelling because the results of the model 
are dependent on this selection. The selection of reference 
spectra must be a representative of surface composition and should be based on 
some prior knowledge of the surface composition. Prior knowledge of 
physical properties is also required for selection of end members.  If the 
end members selection is incorrect in a physical sense, then the 
fractional abundance output results of the model are also potentially 
meaningless. Thus, the end members should be selected 
through an educated guess based on laboratory studies. We have chosen  the 
reference 
spectra based on our knowledge of primitive meteorite composition. 

We selected a total of 30 end members that were supplied to the mixing 
algorithm from a larger spectral library obtained from the University of 
Winnipeg Planetary Spectrophotometer Facility (PSF) in the wavelength range 0.8 to 2.45~$\mu$m. Details of the PSF are available on the PSF web site 
(\url{http://psf.uwinnipeg.ca}). Table~\ref{tab_lab} lists the 
samples selected that broadly includes 
magnetites, clays and silicates. The `dark' minerals which are expected to be 
found on irregular satellites based on previous 
studies \cite[e.g.][]{Adams1974a,1978Adams,1986Cloutis,Clark1990, 
1991Cloutis.Gaffey, 2004cloutis,2007Klima} are specifically included in our 
list of end members.  Metal, magnetites, and amorphous carbon are 
variably effective at suppressing mafic silicate absorption bands 
\citep{Cloutis1990}. The carbon in the list is a synthetic amorphous carbon 
(LCA101,$<$0.021~$\mu$m grain size). The iron oxide group is represented by 
natural magnetite   MAG101 and MAG102 ($Fe_3O_4$). The presence of 
magnetites is indicative of past aqueous alteration in any planetary body \cite[e.g.][]{1980Bunch,1999Hua,2007Rubin}.
Troilite, a dark and spectrally red absorber, is 
also added to the suite of end members because it has been detected 
in minor amounts in ordinary chondrite meteorites and up to 60\% in Fe-Ni 
meteorites \citep{1992Britt}. Silicates that are expected 
to be present initially have also been included in our
end members. Clay minerals have also been included because they are 
formed as alteration products of mafic minerals only in the presence of water 
at the surface or in the sub-surface of C-type asteroids \citep{2002bookRivkin}. 
All sample spectra correspond to bidirectional reflectance 
acquired with a $30^{\circ}$ incidence angle, a $0^{\circ}$ emergence angle, 
and grain size $<$45~$\mu$m. The only exception is the sample LCA101, which has 
a grain size $<$0.021~$\mu$m and whose spectra were obtained with a 
configuration of $0^{\circ}$ incidence and $15^{\circ}$ emergence angles.

\subsection{Application to laboratory mixtures}

%%%%%%%%%%%%%%%%%Figure%%%%%%%%%%%%%%%%%%%%%%%%
\begin{figure*}
\centering

\includegraphics[width=\textwidth]{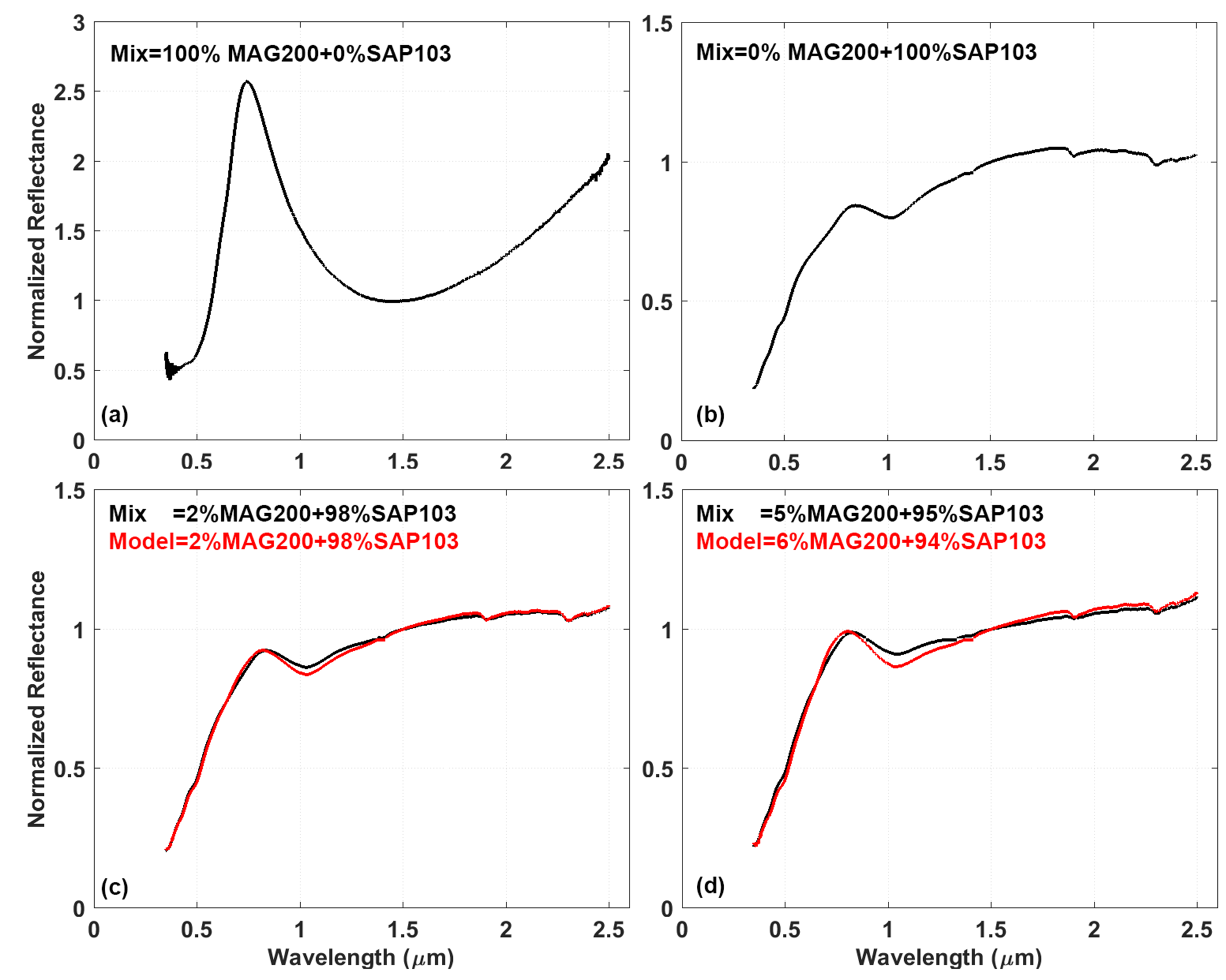}

\caption{Linear unmixing applied to mixtures of magnetite and saponite 
(samples MAG200 and SAP103). (a) and (b) show the end-member spectra. (c) 
and (d) are the results of a linear unmixing model compared with the 
laboratory spectra.  The $R^2$ value is $>0.97$ for both the combinations.}
\label{fig_mag_model}
\end{figure*}
%%%%%%%%%%%%%%%%%Figure%%%%%%%%%%%%%%%%%%%%%%%%
%%%%%%%%%%%%%%%%%Figure%%%%%%%%%%%%%%%%%%%%%%%%
\begin{figure*}
\centering

\includegraphics[width=\textwidth]{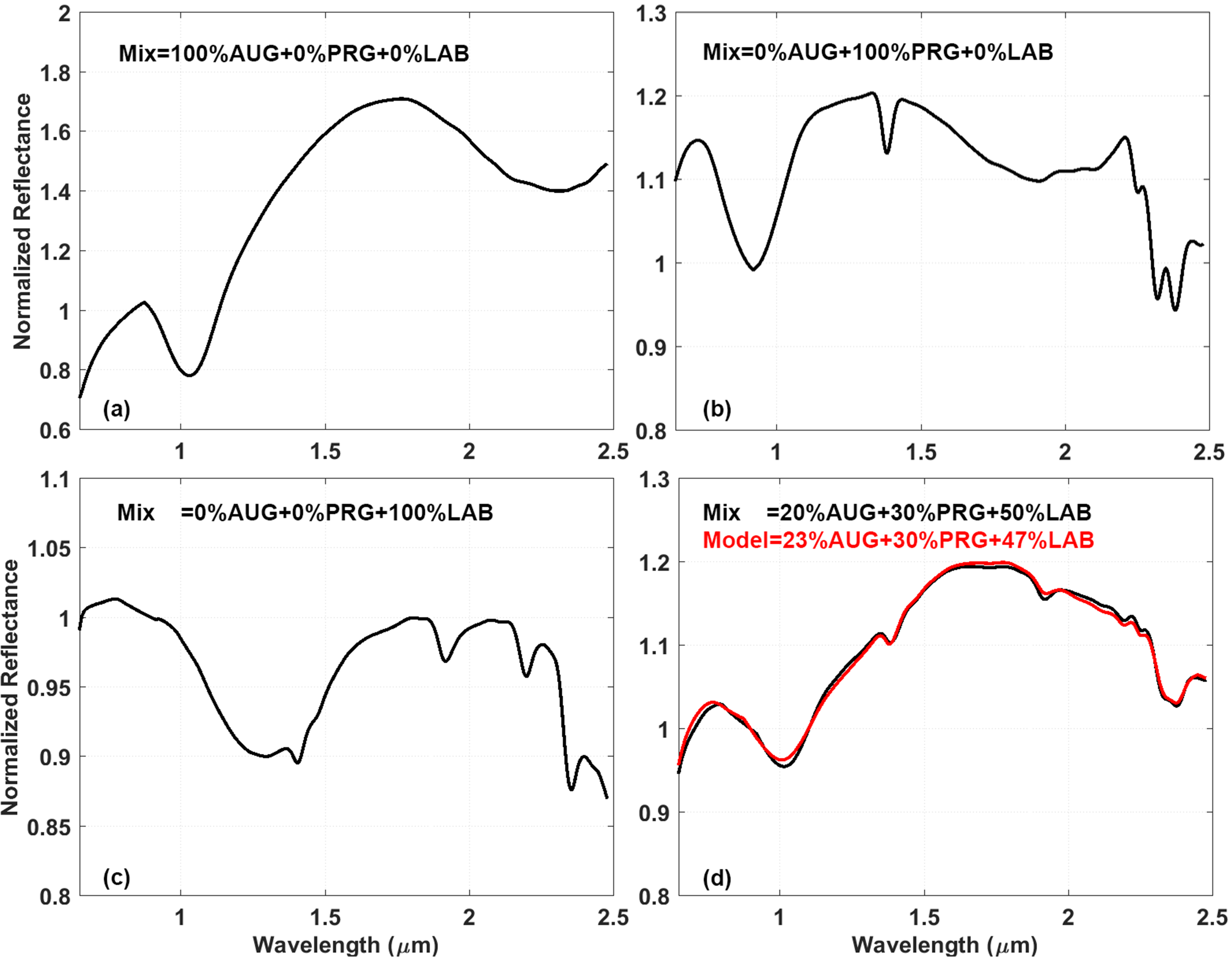}

\caption{Linear unmixing applied to mixtures of Augite, Pargasite and 
Labradorite. (a),(b), and (c) are the end members and (d) is the result of 
linear unmixing model compared with the laboratory measured spectrum.  The 
$R^2$ value is $>0.97$. The mineral mixture's reflectance 
spectrum is normalised at 0.9~$\mu$m.}
\label{fig_pag_model}
\end{figure*}
%%%%%%%%%%%%%%%%%Figure%%%%%%%%%%%%%%%%%%%%%%%%

We present two case studies where we attempt to constrain the possible mineral 
abundances using spectra of two and three mineral mixtures. Our aim is to 
validate the model on known composition and estimate the model accuracy. The  
results are then compared with the measured values by considering $R^2$ and 
$\chi^2$ values. Our model is based on normalised values and cannot distinguish 
albedo differences. We selected two sets of mixtures,  one is a two-mineral 
mixture of  carbonaceous chondrite  simulants and another a three-mineral 
mixture of Augite, Pargasite, and Labradorite. 
Figures~\ref{fig_mag_model} and \ref{fig_pag_model}
summarise the modelling results obtained using the known mineral mixture 
compositions.  

In the case of Jupiter's irregular satellites, we are interested in modelling 
minerals with low albedo, that is, iron oxides, amorphous carbon, clays and 
phyllosilicates. Thus, we selected intimate mixtures of carbonaceous 
chondrite simulants (intimate mixtures of magnetite and saponite) from the PSF 
spectral database; the modelling results are shown for two cases  in 
Fig.~\ref{fig_mag_model}. The $R^2$ values are $>0.97$ for all cases. We observed that the model's absorption band depth systematically 
increases with increasing amounts of magnetite. The maximum deviation in model 
estimation is observed for a 20:80 mixing ratio of MAG200 and SAP103. The model 
in this case estimated a 18:82 mixing ratio. We tested our model for a 
few more two-mineral mixtures and found that the model follows systematic 
trends. 
However, the absolute values may  differ by 10-15~\% in some cases. We 
considered one extreme example in this scenario by modelling intimate mixtures 
of synthetic brucite (OOH032) and LCA101. The absolute coefficients are 
overestimated by 20~\% for intimate mixtures containing  more than 
0.5~\% brucite, despite a good cross-correlation ($>$85\%). However, the 
$R^2$ values significantly decreased ($<$0.65). Thus, we consider a model 
result reliable only if the $R^2$ value is $>$0.70. 

In another case study, we selected a mixture of three known minerals. We 
could only find a limited number of three-mineral mixtures on public spectral 
databases and none of them represent carbonaceous chondrite composition. 
Therefore, we selected the laboratory-based measurements reported recently by 
\cite{2017Rommel}. We do not have a series of laboratory measurements 
in this case by gradually changing the mixing ratios of the end-members. 
Therefore, we considered one available combination of end-members as shown in  
Fig.~\ref{fig_pag_model}. 
The model result is in good agreement with the measured spectrum with $R^2$=0.9.  We also tested 
our model on a few more mixtures of three minerals from the PSF spectral database 
and found that we can identify the increasing/decreasing trends of the minerals 
in three-mineral mixtures; however, the absolute quantification of the 
mineral percentages may deviate by more than 10~\% in some cases.  

\section{Results} \label{sec:result}

\subsection{Himalia spectral comparison with C-type asteroids}

%%%%%%%%%%%%%%%%%Figure%%%%%%%%%%%%%%%%%%%%%%%%
\begin{figure}
\centering

\includegraphics[width=0.6\textwidth]{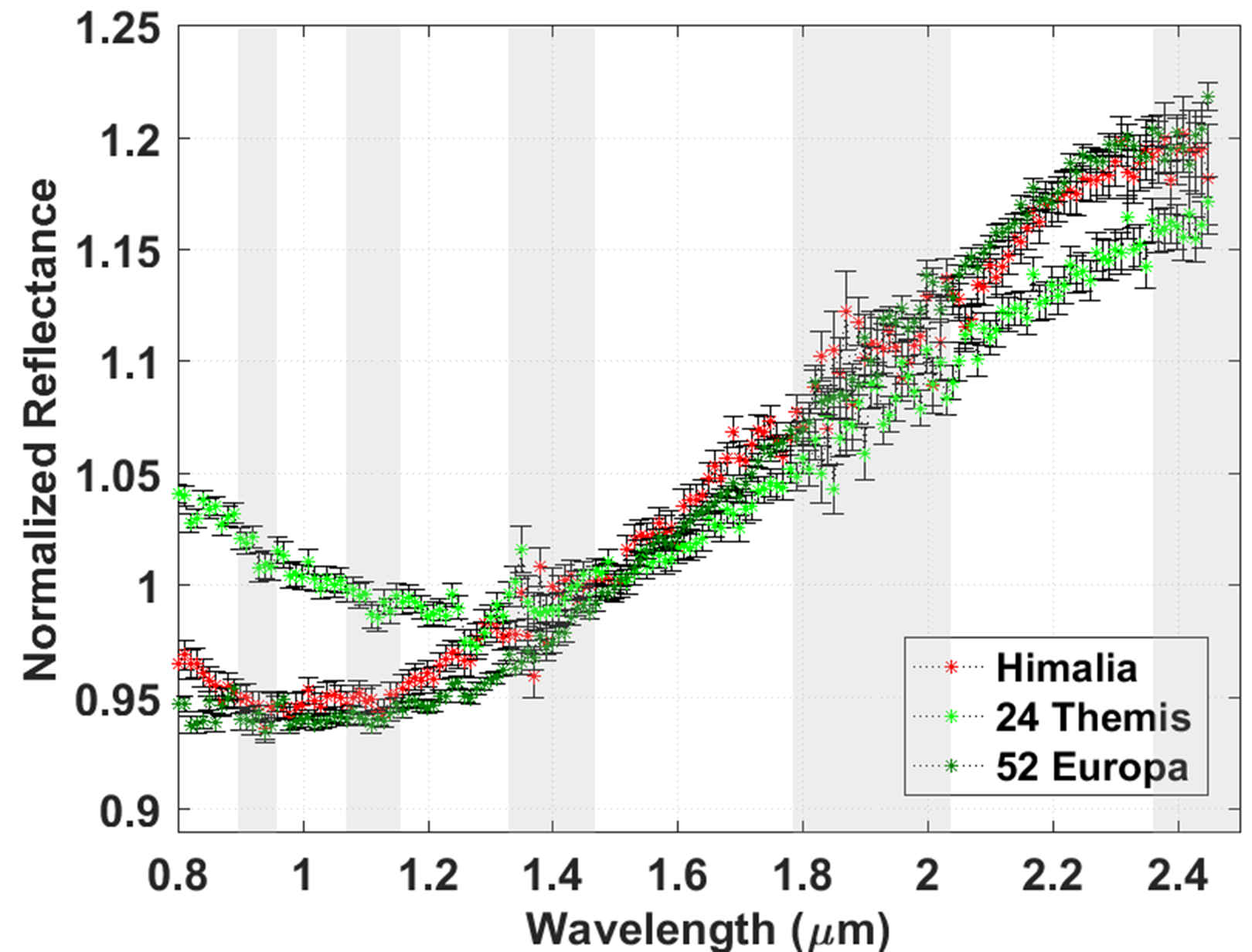}

\caption{The averaged spectrum of Himalia compared to the 
main belt C-type asteroids (52)~Europa  and (24)~Themis. The absorption band 
shapes of Himalia closely match (52)~Europa. All observations
are normalised to unity at 1.5~$\mu$m. The grey bands indicate the location of 
the telluric bands}
\label{fig_HTE}
\end{figure}
%%%%%%%%%%%%%%%%%Figure%%%%%%%%%%%%%%%%%%%%%%%%

The average spectrum of both Himalia and Elara exhibits a broad
absorption band centred around 1.2~$\mu$m which is also 
present in some CM2 carbonaceous chondrites that imply the presence of ferric 
iron with little or no ferrous iron present on the surface 
\citep{2011Cloutis,2011Cloutis_CI}. However, \cite{2014Brown} did not find 
a one to one correspondence between the spectral features of  
Himalia and  a limited set of CM carbonaceous chondrites. Instead, 
\cite{2014Brown} found spectral similarities 
between Himalia and (52)~Europa in the wavelength range 2.2 to 3.8~$\mu$m. We 
compared  a normalised average spectrum of Himalia in 
Fig.~\ref{fig_HTE} with the NIR spectrum of large asteroids, 
(52)~Europa and (24)~Themis.  (52)~Europa and  (24)~Themis are  classified as C 
type under the Bus-DeMeo taxonomic system \citep{2009DeMeo}. The absorption 
band shape and depth of Himalia show a closer resemblance to  
(52)~Europa than to (24)~Themis. Our results from Fig.~\ref{fig_HTE} 
support \cite{2014Brown} findings that spectral characteristics of Himalia fit 
into one of the four spectral categories identified by \cite{2012Takir}
for dark asteroids in the outer asteroid belt. The surface composition of the 
asteroids with Europa-like spectra is unknown and the only way to constrain 
their surface composition is by comparing them with the end members of known 
mineral compositions.  The average spectrum of 
Elara, however, closely resembles that of Himalia up to 1.4~$\mu$m, but with greater 
scattering.

\subsection{Modelling Himalia spectra}

%%%%%%%%%%%%%%%%%Figure%%%%%%%%%%%%%%%%%%%%%%%%
\begin{figure*}
\centering

\includegraphics[width=\textwidth]{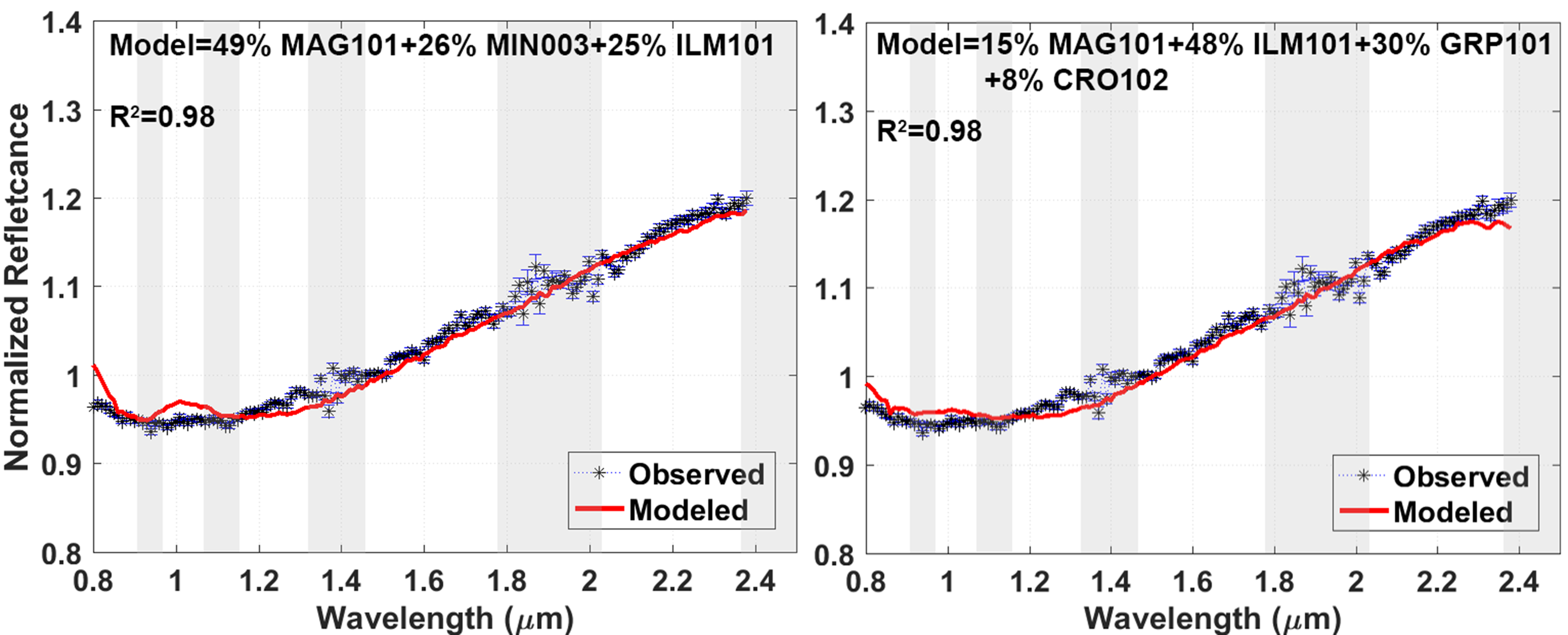}

\caption{The results of linear unmixing applied to Himalia's averaged spectrum 
using three (left) and four (right) end members. Out of ten runs, the best case 
with the lowest $\chi^2$  and maximum $R^2$ is shown here. The grey bands 
indicate the location of the telluric bands.}
\label{fig_H_model}
\end{figure*}
%%%%%%%%%%%%%%%%%Figure%%%%%%%%%%%%%%%%%%%%%%%%

Fig.~\ref{fig_H_model} shows the results obtained using a linear unmixing 
model for Himalia. The best match is a mixture of iron oxides (magnetite and 
ilmenite) with a ferric phyllosilicate (minnesotaite) considering three 
end members in the linear unmixing algorithm. The model mainly deviates 
from the observations in the shorter wavelength range from 0.8 to 1.4~$\mu$m. 
Therefore, to further improve the model we increased the end members 
from three to four. The best fit result using four end members is also 
shown in Fig.~\ref{fig_H_model} and  gives a 
better match for the full wavelength region with no significant change in the  
$R^2$ values. In this case, minnesotaite has been replaced by a combination of 
carbon (graphite) and a different ferric phyllosilicate, cronstedtite, which is 
also present in CM2 carbonaceous chondrites. The model does not match the 
observations in the wavelength range 1.2-1.4~$\mu$m even if we further 
increase the end members from four to five. 

\cite{2000Jarvis} suggested the possible  presence of phyllosilicates based on 
the presence of 0.7~$\mu$m absorption feature and classified the parent body of 
Himalia as large main belt asteroid, a subclass of C, and proposed that it 
belongs to the Nysa asteroid family. Our modelling 
results support the findings of \cite{2000Jarvis}  by estimating phyllosilicates using three and
four end members models as shown in Fig.~\ref{fig_H_model}.

\subsection{Modelling Elara Spectra}

%%%%%%%%%%%%%%%%%Figure%%%%%%%%%%%%%%%%%%%%%%%%
\begin{figure}
\centering

\includegraphics[width=0.7\textwidth]{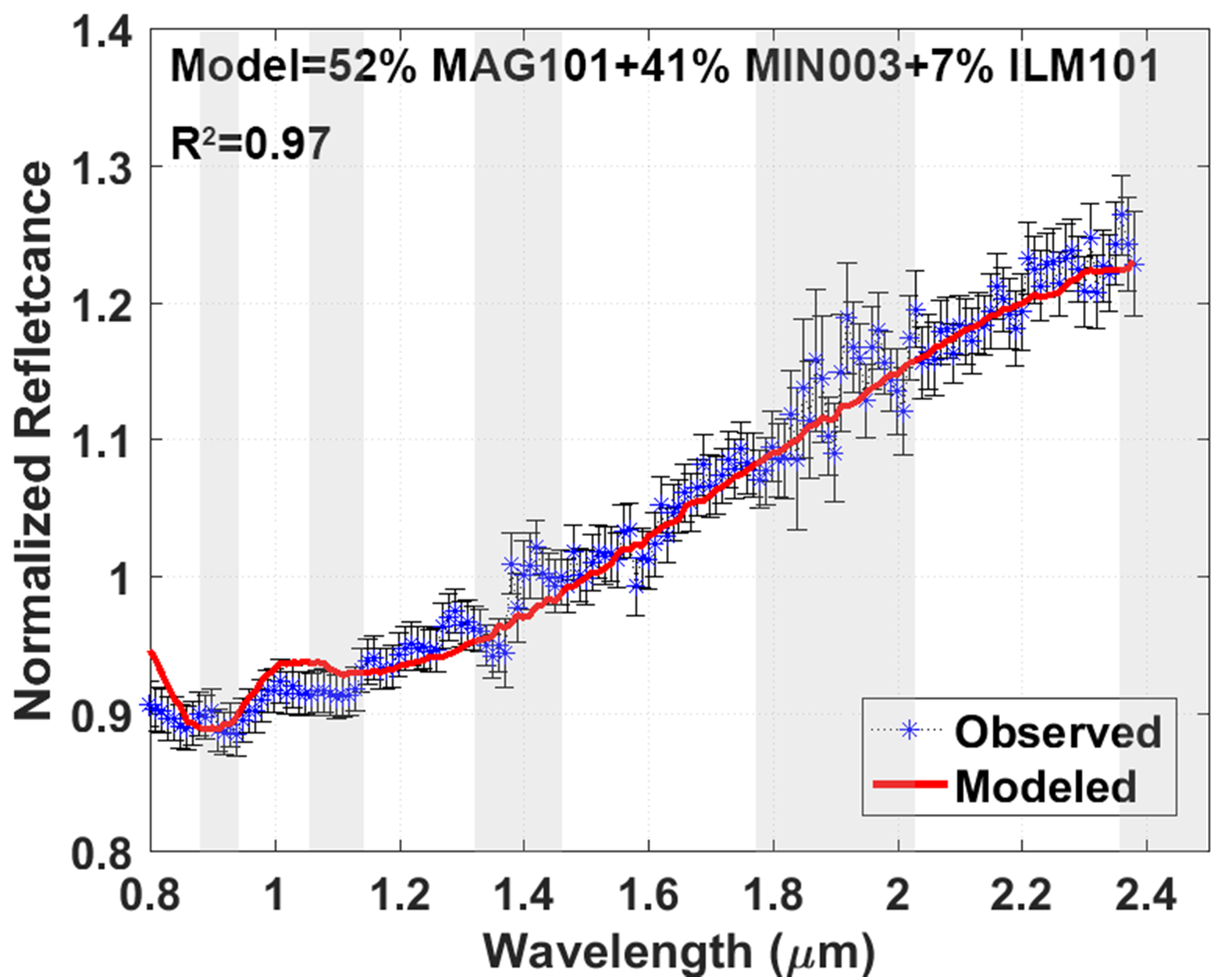}

\caption{Results of the linear unmixing model applied to the averaged near-IR 
spectrum of Elara. The mineral composition of Elara is similar to Himalia with 
exactly the same mineral combination found but with different mixing 
coefficients. The grey bands indicate the location of the telluric bands.}
\label{fig_E_model}
\end{figure}
%%%%%%%%%%%%%%%%%Figure%%%%%%%%%%%%%%%%%%%%%%%%
Elara  has been suggested as having a 
similar surface composition as Himalia because it belongs to the Himalia 
spectral family \citep{1991Luu,2001Rettig,2003Grav,2004Grav}. 
Figure~\ref{fig_E_model} shows the modelling results for Elara where we 
find the same set of minerals for Elara as we found for Himalia,  indicating a 
composition dominated by iron oxides and phyllosilicates. We also checked 
the model result by increasing the end members to four in 
order to improve the fit at shorter wavelength region 
($<$1.4~$\mu$m) but 
did not find any significant improvement in the fit. This analysis confirms 
that Himalia and Elara plausibly belong to a single parent body. However, we 
note that Elara has redder slope compared to Himalia, which may also indicate 
different surface alteration processes. 

\subsection{Modelling Carme Spectra}

%%%%%%%%%%%%%%%%%Figure%%%%%%%%%%%%%%%%%%%%%%%%
\begin{figure}
\centering

\includegraphics[width=0.7\textwidth]{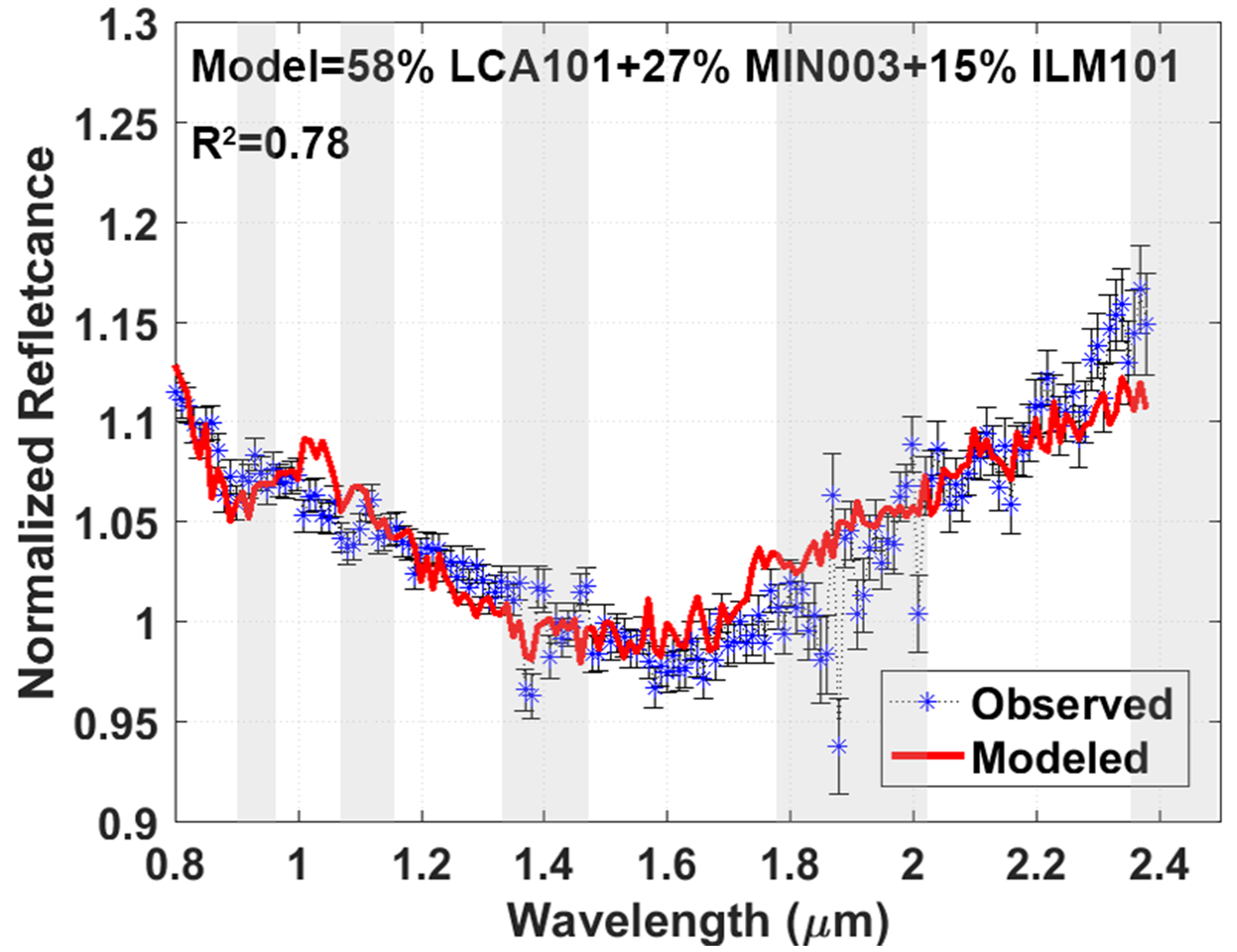}

\caption{Results of the linear unmixing model applied to the NIR spectrum 
of Carme. The model provides the best match to the observed spectrum with  
$R^2$ = 0.78. The grey bands indicate the location of the telluric bands.}
\label{fig_C_model}
\end{figure}
%%%%%%%%%%%%%%%%%Figure%%%%%%%%%%%%%%%%%%%%%%%%
Figure~\ref{fig_C_model} shows the best spectral fit 
obtained using our model for Carme. Carme's 
normalised reflectance spectrum  has  a broad  
absorption band centered around 1.6~$\mu$m. We find that the Carme spectrum has
several interesting spectral features in the NIR 
wavelength range not previously detected. The overall spectral 
shape and absorption features of Carme are unique and distinguish
it from other satellites. Our model suggests that the 
main constituents of the 
surface are black carbon, minnesotaite, and ilmenite. The model 
failed to match 
the telescopic spectrum  at the wavelength range 0.9 to 1.1~$\mu$m  even when 
we increased the number of end members to four. 
 
\cite{1991Luu} and \cite{1984Tholen} reported the visible spectrum of Carme 
as consistent with D-type asteroids due to red slope from 0.4 to 0.7~$\mu$m. 
However, our near-IR observations are significantly different from 
visible wavelength observations  of 
\cite{1991Luu} and \cite{1984Tholen}. The multicolour photometry 
observations in the 
ultraviolet (UV) wavelength region   by \cite{1984Tholen}  found a strong 
upturn around 0.3~$\mu$m that could be due to 
possible low-level cometary activity at Carme. This characteristic of Carme is 
unique and distinguishes it from other irregular satellites. As we do not cover 
observations in UV or the visible wavelength regions we could not confirm these 
findings.

\section{Discussion} \label{sec:disc}
Our modelling and spectral shape comparison results in the NIR wavelength range for Himalia and Elara suggest similar surface composition as also demonstrated by 
several previous studies 
\cite[e.g.][]{1977Cruikshank,1980Degewij,1984Tholen,1991Luu,2000Brown,
2001Rettig, 2003Grav}. However, our Carme observations in the near-IR wavelengths distinguish this study  
from previous visible observations \citep{1984Tholen,1991Luu}. 
Our analysis of these three irregular satellites  provides information on the major mineral contributors by 
comparing observations with laboratory-based studies.  The modelling 
approach proposed in this study, however, has certain limitations.  One of the main 
limitations is that the model depends  on the defined initial conditions such as selection of end members, grain 
sizes, viewing geometry, and so on. We compared the laboratory measured 
reflectance spectra with known physical conditions to the spectra observed from 
distant objects at surface temperature much lower than ambient
temperature on Earth.  Another limitation is that we could not consider 
wavelengths shorter than 0.8~$\mu$m due to poor signal to noise ratio, 
limiting our ability to detect the 0.7~$\mu$m absorption feature. Our telescopic 
observations do not cover the 3~$\mu$m region, which would 
help to improve our understanding of the primordial parent body of these 
irregular satellites. The 3~$\mu$m region includes several features 
of water, ice, and hydroxyl-bearing minerals, such as the 3.1~$\mu$m absorption 
feature due to the water molecule in ice, and the 2.9~$\mu$m absorption feature 
due to the O-H bond in hydroxyls \citep{1984Aines,1990Jones}. \cite{1999Howell} and 
\cite{2002bookRivkin} concluded that the presence of the 0.7~$\mu$m absorption 
feature in a spectrum always indicates the presence of the 3.0~$\mu$m feature, 
although the reverse is not always true. 

 A comparison of NIR spectra of Himalia  with the 
two large main belt asteroids (24) Themis and (52) Europa suggests that the 
parent body(ies) of  the Himalia family might have been derived 
from the main asteroid belt. This conclusion has been advanced by several 
studies \cite[e.g.][]{1977Cruikshank,1980Degewij,1984Tholen,1991Luu,2000Brown,
2001Rettig, 2003Grav}. The origin of the parent body(ies) of irregular 
satellites is still an open question and a more definitive answer requires 
observations of several irregular satellites from each dynamical family with 
higher signal-to-noise ratio at full wavelength coverage from 0.6 to 
3~$\mu$m.   Planetesimals from different 
regions of the solar system are expected to show different spectral signatures 
in this wavelength range, and identifying these absorption 
features is the  key to resolving the question of the origin of the parent 
bodies of the irregular satellites.  

The low albedo values of Himalia, Elara, and Carme \citep{2015Grav} suggest 
that a darkening agent is mixed in the surface material and also that the 
absorption features we observe are actually very intense in order to appear in 
reflectance spectra with low albedos. The darkening agent was identified as 
magnetite for Himalia and Elara, and as amorphous carbon for Carme along with 
the presence of phyllosilicates.  The coexistence of phyllosilicates and iron 
oxides is due to aqueous alteration based on the observed mineralogy of 
carbonaceous chondrites.  Moreover, the evidence for coexistence of 
phyllosilicates and iron oxides in the same planetary source already exists, 
for example, CI1, CR2, and CM2 carbonaceous chondrites \cite[e.g.][]{1994Vilas}. The 
detection of magnetite along with phyllosilicate  indicates that Himalia and 
Elara may belong to the C-type asteroids from the main belt as also reported by 
previous studies. 
\section{Conclusions} \label{sec:con}
Surface composition of the Jovian irregular satellites helps us
constrain the nature and origin of their parent bodies. Here, we
identified major mineral contributors by applying a spectral unmixing
model to VNIR reflectance spectra of Himalia, Elara, and
Carme. We first evaluated the model performance on a number
of intimately mixed mineral powders of known grain sizes and
compositions. We discussed the limitations of our model in order
to ultimately be able to estimate optically dominant minerals
on very low albedo objects.

The spectral unmixing model solution has been considered stable 
for Himalia, Elara, and Carme considering similar composition from ten
automatic model runs with R$^2$ > 0.75. The normalised reflectance
spectra of Himalia and Elara are spectrally similar, suggesting
similar surface composition that might have been derived 
from the main asteroid belt. Carme is unique in its spectral shape and absorption features.
The reflectance spectra of these three objects are dominated by iron oxides and phyllosilicates,
an important indicator of past aqueous alteration of their parent body(s). The spectral unmixing model
revealed that the most dominant mineral is magnetite for Himalia and Elara, and amorphous
carbon for Carme.

The modelling results of Himalia and Elara suggest a very reduced environment 
with a negligible amount of  Fe$^3+$ present on these bodies. The detection of 
water or minerals which can only be formed in the presence of water is important in 
order to understand the nature of the parent body(ies) of the irregular 
satellites. This information also sheds light on the processes that occurred during 
the early stages of solar system history. Although we do not cover the 3~$\mu$m 
region for detecting hydration features, we found phyllosilicates present in 
each case using the modelling approach.We plan to observe additional Jovian 
irregular satellites both from prograde and retrograde groups covering the full 
wavelength range between 0.6 and 3~$\mu$m in future in order to quantify the 
extent of aqueous alteration from these bodies.

\begin{acknowledgements}
The authors wish to thank one anonymous reviewer and Guneshwar 
Thangjam for their helpful reviews of the manuscript. MB is funded by the 
Indian Space Research Organization through its post doctoral 
program.  We thank Prof. Christian W\"ohler, TU Dortmund for providing 
three mineral mixture reflectance spectra. We would like to  thank Juan A. 
Sanchez for his comments to improve the manuscript. Work by VR  was funded by 
NASA Planetary Geology and Geophysics grants NNX14AN05G and NNX14AN35G. EAC 
thanks the Canada Foundation for Innovation, the Manitoba Research Innovations 
Fund, the Canadian Space Agency, the Natural Sciences and Engineering Research 
Council of Canada, and the University of Winnipeg for supporting the 
establishment and operation of the Planetary Spectrophotometer Facility.
\end{acknowledgements}

   \bibliographystyle{aa} % style aa.bst
   \bibliography{reference} % your references Yourfile.bib

\begin{thebibliography}{65}
\expandafter\ifx\csname natexlab\endcsname\relax\def\natexlab#1{#1}\fi

\bibitem[{{Adams}(1974)}]{Adams1974a}
{Adams}, J.~B. 1974, \jgr, 79, 4829

\bibitem[{{Adams} \& {Goullaud}(1978)}]{1978Adams}
{Adams}, J.~B. \& {Goullaud}, L.~H. 1978, in Lunar~Planet.~Sci.~Conf., Vol.~9,
  1--3

\bibitem[{{Adams} {et~al.}(1986){Adams}, {Smith}, \& {Johnson}}]{1986Adams}
{Adams}, J.~B., {Smith}, M.~O., \& {Johnson}, P.~E. 1986, \jgr, 91, 8098

\bibitem[{{Aines} \& {Rossman}(1984)}]{1984Aines}
{Aines}, R.~D. \& {Rossman}, G.~R. 1984, \jgr, 89, 4059

\bibitem[{{Bailey}(1971{\natexlab{a}})}]{1971Bailey}
{Bailey}, J.~M. 1971{\natexlab{a}}, Science, 173, 812

\bibitem[{{Bailey}(1971{\natexlab{b}})}]{1971BaileyJGR}
{Bailey}, J.~M. 1971{\natexlab{b}}, \jgr, 76, 7827

\bibitem[{{Britt} {et~al.}(1992){Britt}, {Tholen}, {Bell}, \&
  {Pieters}}]{1992Britt}
{Britt}, D.~T., {Tholen}, D.~J., {Bell}, J.~F., \& {Pieters}, C.~M. 1992,
  \icarus, 99, 153

\bibitem[{{Brown}(2000)}]{2000Brown}
{Brown}, M.~E. 2000, \aj, 119, 977

\bibitem[{{Brown} \& {Rhoden}(2014)}]{2014Brown}
{Brown}, M.~E. \& {Rhoden}, A.~R. 2014, \apjl, 793, L44

\bibitem[{{Brown} {et~al.}(2003){Brown}, {Baines}, {Bellucci}, {Bibring},
  {Buratti}, {Capaccioni}, {Cerroni}, {Clark}, {Coradini}, {Cruikshank},
  {Drossart}, {Formisano}, {Jaumann}, {Langevin}, {Matson}, {McCord},
  {Mennella}, {Nelson}, {Nicholson}, {Sicardy}, {Sotin}, {Amici},
  {Chamberlain}, {Filacchione}, {Hansen}, {Hibbitts}, \&
  {Showalter}}]{2003Brown}
{Brown}, R.~H., {Baines}, K.~H., {Bellucci}, G., {et~al.} 2003, \icarus, 164,
  461

\bibitem[{{Bunch} \& {Chang}(1980)}]{1980Bunch}
{Bunch}, T.~E. \& {Chang}, S. 1980, \gca, 44, 1543

\bibitem[{{Chamberlain} \& {Brown}(2004)}]{2004Chamberlain}
{Chamberlain}, M.~A. \& {Brown}, R.~H. 2004, \icarus, 172, 163

\bibitem[{{Clark} {et~al.}(1990){Clark}, {King}, {Klejwa}, {Swayze}, \&
  {Vergo}}]{Clark1990}
{Clark}, R.~N., {King}, T.~V.~V., {Klejwa}, M., {Swayze}, G.~A., \& {Vergo}, N.
  1990, \jgr, 95, 12653

\bibitem[{{Cloutis} \& {Gaffey}(1991)}]{1991Cloutis.Gaffey}
{Cloutis}, E.~A. \& {Gaffey}, M.~J. 1991, \jgr, 96, 22809

\bibitem[{{Cloutis} {et~al.}(1986){Cloutis}, {Gaffey}, {Jackowski}, \&
  {Reed}}]{1986Cloutis}
{Cloutis}, E.~A., {Gaffey}, M.~J., {Jackowski}, T.~L., \& {Reed}, K.~L. 1986,
  \jgr, 91, 11641

\bibitem[{Cloutis {et~al.}(1990)Cloutis, Gaffey, Smith, \&
  Lambert}]{Cloutis1990}
Cloutis, E.~A., Gaffey, M.~J., Smith, D. G.~W., \& Lambert, R. S.~J. 1990,
  Journal of Geophysical Research: Solid Earth, 95, 8323

\bibitem[{{Cloutis} {et~al.}(2011{\natexlab{a}}){Cloutis}, {Hiroi}, {Gaffey},
  {Alexander}, \& {Mann}}]{2011Cloutis_CI}
{Cloutis}, E.~A., {Hiroi}, T., {Gaffey}, M.~J., {Alexander}, C.~M.~O.~., \&
  {Mann}, P. 2011{\natexlab{a}}, \icarus, 212, 180

\bibitem[{{Cloutis} {et~al.}(2011{\natexlab{b}}){Cloutis}, {Hudon}, {Hiroi},
  {Gaffey}, \& {Mann}}]{2011Cloutis}
{Cloutis}, E.~A., {Hudon}, P., {Hiroi}, T., {Gaffey}, M.~J., \& {Mann}, P.
  2011{\natexlab{b}}, \icarus, 216, 309

\bibitem[{{Cloutis} {et~al.}(2004){Cloutis}, {Sunshine}, \&
  {Morris}}]{2004cloutis}
{Cloutis}, E.~A., {Sunshine}, J.~M., \& {Morris}, R.~V. 2004, Meteoritics and
  Planetary Science, 39, 545

\bibitem[{{Colombo} \& {Franklin}(1971)}]{1971Colombo}
{Colombo}, G. \& {Franklin}, F.~A. 1971, \icarus, 15, 186

\bibitem[{{Combe} {et~al.}(2008){Combe}, {Le Mou{\'e}lic}, {Sotin}, {Gendrin},
  {Mustard}, {Le Deit}, {Launeau}, {Bibring}, {Gondet}, {Langevin}, {Pinet}, \&
  {OMEGA Science Team}}]{2008Combe}
{Combe}, J.-P., {Le Mou{\'e}lic}, S., {Sotin}, C., {et~al.} 2008, \planss, 56,
  951

\bibitem[{{Cruikshank}(1977)}]{1977Cruikshank}
{Cruikshank}, D.~P. 1977, \icarus, 30, 224

\bibitem[{{Cushing} {et~al.}(2004){Cushing}, {Vacca}, \&
  {Rayner}}]{2004Cushing}
{Cushing}, M.~C., {Vacca}, W.~D., \& {Rayner}, J.~T. 2004, \pasp, 116, 362

\bibitem[{{Degewij} {et~al.}(1980){Degewij}, {Cruikshank}, \&
  {Hartmann}}]{1980Degewij}
{Degewij}, J., {Cruikshank}, D.~P., \& {Hartmann}, W.~K. 1980, \icarus, 44, 541

\bibitem[{{DeMeo} {et~al.}(2009){DeMeo}, {Binzel}, {Slivan}, \&
  {Bus}}]{2009DeMeo}
{DeMeo}, F.~E., {Binzel}, R.~P., {Slivan}, S.~M., \& {Bus}, S.~J. 2009,
  \icarus, 202, 160

\bibitem[{{Grav} {et~al.}(2015){Grav}, {Bauer}, {Mainzer}, {Masiero}, {Nugent},
  {Cutri}, {Sonnett}, \& {Kramer}}]{2015Grav}
{Grav}, T., {Bauer}, J.~M., {Mainzer}, A.~K., {et~al.} 2015, \apj, 809, 3

\bibitem[{{Grav} \& {Holman}(2004)}]{2004Grav}
{Grav}, T. \& {Holman}, M.~J. 2004, \apjl, 605, L141

\bibitem[{{Grav} {et~al.}(2003){Grav}, {Holman}, {Gladman}, \&
  {Aksnes}}]{2003Grav}
{Grav}, T., {Holman}, M.~J., {Gladman}, B.~J., \& {Aksnes}, K. 2003, \icarus,
  166, 33

\bibitem[{{Hapke}(1981)}]{1981Hapke}
{Hapke}, B. 1981, \jgr, 86, 4571

\bibitem[{{Hapke}(2002)}]{2002Hapke}
{Hapke}, B. 2002, \icarus, 157, 523

\bibitem[{{Hartmann}(1987)}]{1987Hartmann}
{Hartmann}, W.~K. 1987, \icarus, 71, 57

\bibitem[{{Heppenheimer} \& {Porco}(1977)}]{1977Heppen}
{Heppenheimer}, T.~A. \& {Porco}, C. 1977, \icarus, 30, 385

\bibitem[{{Heylen} {et~al.}(2011){Heylen}, {Burazerovic}, \&
  {Scheunders}}]{2011Heylen}
{Heylen}, R., {Burazerovic}, D., \& {Scheunders}, P. 2011, IEEE Transactions on
  Geoscience and Remote Sensing, 49, 4112

\bibitem[{{Horgan} {et~al.}(2014){Horgan}, {Cloutis}, {Mann}, \&
  {Bell}}]{2014Horgan}
{Horgan}, B.~H.~N., {Cloutis}, E.~A., {Mann}, P., \& {Bell}, J.~F. 2014,
  \icarus, 234, 132

\bibitem[{{Howell} {et~al.}(1999){Howell}, {Rivkin}, {Soderberg}, {Vilas}, \&
  {Cochran}}]{1999Howell}
{Howell}, E.~S., {Rivkin}, A.~S., {Soderberg}, A., {Vilas}, F., \& {Cochran},
  A.~L. 1999, in AAS/Division for Planetary Sciences Meeting Abstracts,
  Vol.~31, AAS/Division for Planetary Sciences Meeting Abstracts \#31, 04.01

\bibitem[{{Hua} \& {Buseck}(1999)}]{1999Hua}
{Hua}, X. \& {Buseck}, P.~R. 1999, Meteoritics and Planetary Science, 34, A187

\bibitem[{{Jacobson}(2000)}]{2000Jacobson}
{Jacobson}, R.~A. 2000, \aj, 120, 2679

\bibitem[{{Jarvis} {et~al.}(2000){Jarvis}, {Vilas}, {Larson}, \&
  {Gaffey}}]{2000Jarvis}
{Jarvis}, K.~S., {Vilas}, F., {Larson}, S.~M., \& {Gaffey}, M.~J. 2000,
  \icarus, 145, 445

\bibitem[{{Jewitt} \& {Haghighipour}(2007)}]{2007Jewitt}
{Jewitt}, D. \& {Haghighipour}, N. 2007, \araa, 45, 261

\bibitem[{{Jones} {et~al.}(1990){Jones}, {Lebofsky}, {Lewis}, \&
  {Marley}}]{1990Jones}
{Jones}, T.~D., {Lebofsky}, L.~A., {Lewis}, J.~S., \& {Marley}, M.~S. 1990,
  \icarus, 88, 172

\bibitem[{{Klima} {et~al.}(2007){Klima}, {Pieters}, \& {Dyar}}]{2007Klima}
{Klima}, R.~L., {Pieters}, C.~M., \& {Dyar}, M.~D. 2007, Meteoritics and
  Planetary Science, 42, 235

\bibitem[{{Luu}(1991)}]{1991Luu}
{Luu}, J. 1991, \aj, 102, 1213

\bibitem[{{Marsset} {et~al.}(2016){Marsset}, {Vernazza}, {Birlan}, {DeMeo},
  {Binzel}, {Dumas}, {Milli}, \& {Popescu}}]{2016Marsset}
{Marsset}, M., {Vernazza}, P., {Birlan}, M., {et~al.} 2016, \aap, 586, A15

\bibitem[{{Nesvorn{\'y}} {et~al.}(2003){Nesvorn{\'y}}, {Alvarellos}, {Dones},
  \& {Levison}}]{2003Nesvorn}
{Nesvorn{\'y}}, D., {Alvarellos}, J.~L.~A., {Dones}, L., \& {Levison}, H.~F.
  2003, \aj, 126, 398

\bibitem[{{Nesvorn{\'y}} {et~al.}(2014){Nesvorn{\'y}}, {Vokrouhlick{\'y}}, \&
  {Deienno}}]{2014Nersvorny}
{Nesvorn{\'y}}, D., {Vokrouhlick{\'y}}, D., \& {Deienno}, R. 2014, \apj, 784,
  22

\bibitem[{{Nesvorn{\'y}} {et~al.}(2007){Nesvorn{\'y}}, {Vokrouhlick{\'y}}, \&
  {Morbidelli}}]{2007Nesvorny}
{Nesvorn{\'y}}, D., {Vokrouhlick{\'y}}, D., \& {Morbidelli}, A. 2007, \aj, 133,
  1962

\bibitem[{{Nicholson} {et~al.}(2008){Nicholson}, {Cuk}, {Sheppard}, {Nesvorny},
  \& {Johnson}}]{2008Nicholson}
{Nicholson}, P.~D., {Cuk}, M., {Sheppard}, S.~S., {Nesvorny}, D., \& {Johnson},
  T.~V. 2008, {Irregular Satellites of the Giant Planets}, ed. M.~A. {Barucci},
  H.~{Boehnhardt}, D.~P. {Cruikshank}, A.~{Morbidelli}, \& R.~{Dotson},
  411--424

\bibitem[{{Pollack} {et~al.}(1979){Pollack}, {Burns}, \&
  {Tauber}}]{1979Pollack}
{Pollack}, J.~B., {Burns}, J.~A., \& {Tauber}, M.~E. 1979, \icarus, 37, 587

\bibitem[{{Porco} {et~al.}(2003){Porco}, {West}, {McEwen}, {Del Genio},
  {Ingersoll}, {Thomas}, {Squyres}, {Dones}, {Murray}, {Johnson}, {Burns},
  {Brahic}, {Neukum}, {Veverka}, {Barbara}, {Denk}, {Evans}, {Ferrier},
  {Geissler}, {Helfenstein}, {Roatsch}, {Throop}, {Tiscareno}, \&
  {Vasavada}}]{2003Porco}
{Porco}, C.~C., {West}, R.~A., {McEwen}, A., {et~al.} 2003, Science, 299, 1541

\bibitem[{{Poulet} {et~al.}(2009){Poulet}, {Bibring}, {Langevin}, {Mustard},
  {Mangold}, {Vincendon}, {Gondet}, {Pinet}, {Bardintzeff}, \&
  {Platevoet}}]{2009Poulet}
{Poulet}, F., {Bibring}, J.-P., {Langevin}, Y., {et~al.} 2009, \icarus, 201, 69

\bibitem[{{Ramsey} \& {Christensen}(1998)}]{1998Ramsey}
{Ramsey}, M.~S. \& {Christensen}, P.~R. 1998, \jgr, 103, 577

\bibitem[{{Rettig} {et~al.}(2001){Rettig}, {Walsh}, \&
  {Consolmagno}}]{2001Rettig}
{Rettig}, T.~W., {Walsh}, K., \& {Consolmagno}, G. 2001, \icarus, 154, 313

\bibitem[{{Rivkin} {et~al.}(2002){Rivkin}, {Howell}, {Vilas}, \&
  {Lebofsky}}]{2002bookRivkin}
{Rivkin}, A.~S., {Howell}, E.~S., {Vilas}, F., \& {Lebofsky}, L.~A. 2002,
  {Hydrated Minerals on Asteroids: The Astronomical Record}, ed. W.~F.
  {Bottke}, Jr., A.~{Cellino}, P.~{Paolicchi}, \& R.~P. {Binzel}, 235--253

\bibitem[{{Rommel} {et~al.}(2017){Rommel}, {Grumpe}, {Felder}, {W{\"o}hler},
  {Mall}, \& {Kronz}}]{2017Rommel}
{Rommel}, D., {Grumpe}, A., {Felder}, M.~P., {et~al.} 2017, \icarus, 284, 126

\bibitem[{{Rubin} {et~al.}(2007){Rubin}, {Trigo-Rodr{\'{\i}}guez}, {Huber}, \&
  {Wasson}}]{2007Rubin}
{Rubin}, A.~E., {Trigo-Rodr{\'{\i}}guez}, J.~M., {Huber}, H., \& {Wasson},
  J.~T. 2007, \gca, 71, 2361

\bibitem[{{Sanchez} {et~al.}(2013){Sanchez}, {Michelsen}, {Reddy}, \&
  {Nathues}}]{2013Sanchez}
{Sanchez}, J.~A., {Michelsen}, R., {Reddy}, V., \& {Nathues}, A. 2013, \icarus,
  225, 131

\bibitem[{{Sanchez} {et~al.}(2015){Sanchez}, {Reddy}, {Dykhuis}, {Lindsay}, \&
  {Le Corre}}]{2015Sanchez}
{Sanchez}, J.~A., {Reddy}, V., {Dykhuis}, M., {Lindsay}, S., \& {Le Corre}, L.
  2015, \apj, 808, 93

\bibitem[{{Sheppard} \& {Jewitt}(2003)}]{2003sheppard}
{Sheppard}, S.~S. \& {Jewitt}, D.~C. 2003, \nat, 423, 261

\bibitem[{{Shkuratov} {et~al.}(1999){Shkuratov}, {Kreslavsky}, {Ovcharenko},
  {Stankevich}, {Zubko}, {Pieters}, \& {Arnold}}]{Shkuratov1999a}
{Shkuratov}, Y.~G., {Kreslavsky}, M.~A., {Ovcharenko}, A.~A., {et~al.} 1999,
  \icarus, 141, 132

\bibitem[{{Sunshine} {et~al.}(1990){Sunshine}, {Pieters}, \&
  {Pratt}}]{1990Sunshine}
{Sunshine}, J.~M., {Pieters}, C.~M., \& {Pratt}, S.~F. 1990, \jgr, 95, 6955

\bibitem[{{Takir} \& {Emery}(2012)}]{2012Takir}
{Takir}, D. \& {Emery}, J.~P. 2012, \icarus, 219, 641

\bibitem[{{Tholen} \& {Zellner}(1984)}]{1984Tholen}
{Tholen}, D.~J. \& {Zellner}, B. 1984, \icarus, 58, 246

\bibitem[{{van der Meer}(2000)}]{2000van}
{van der Meer}, F. 2000, International Journal of Remote Sensing, 21, 3179

\bibitem[{{Vilas} {et~al.}(1994){Vilas}, {Jarvis}, \& {Gaffey}}]{1994Vilas}
{Vilas}, F., {Jarvis}, K.~S., \& {Gaffey}, M.~J. 1994, \icarus, 109, 274

\bibitem[{{Zambon} {et~al.}(2016){Zambon}, {Tosi}, {Carli}, {De Sanctis},
  {Blewett}, {Palomba}, {Longobardo}, {Frigeri}, {Ammannito}, {Russell}, \&
  {Raymond}}]{2016Zambon}
{Zambon}, F., {Tosi}, F., {Carli}, C., {et~al.} 2016, \icarus, 272, 16

\end{thebibliography}

\end{document}